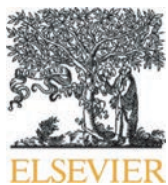
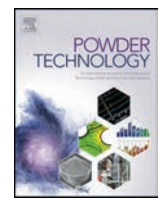

# Dynamics of quasi-static collapse process of a binary granular column

Hongwei Zhu [a,1], Yaodong Feng [a,b,1], Danfeng Lu [a,c,1], Yahya Sandali [a], Bin Li [a], Gang Sun [b,*], Ning Zheng [a,*], Qingfan Shi [a,*]

[a] School of Physics, Beijing Institute of Technology, Beijing 100081, China
[b] Key Laboratory of Soft Matter Physics, Beijing National Laboratory for Condensed Matter Physics, Institute of Physics, Chinese Academy of Sciences, Beijing 100190, China
[c] Optical Center, JiangSu North Huguang Opto-Electronics CO., Ltd, Wuxi 214035, China

A B S T R A C T

The dynamical behavior of the column that made up binary granular beads is investigated systematically by tracking the displacement of particles in the collapse process. An experimental setup is first devised to control the quasi-static collapse of a granular column, and then observe the trajectories of tracer particles by using an industrial camera controlled by the image acquisition program. It is found that there exist two zones in column: a sliding region in which particles are moving in a layered structure; a static region within which particles are stationary. According to this analytical result, a dynamical model is developed to predict the trajectory evolution of particles in the space-time. The calculating result for the trajectories of particles on the selected layers is well consistent with the experimental observation.

© 2018 Published by Elsevier B.V.

## 1. Introduction

The collapse and flow of granular materials have been the focus of an increasing amount of research in recent years, because their properties and behaviors as either static piles or highly mobile flows are fundamental to natural catastrophic events such as landslides [1], avalanches [2], and hazards in civil engineering projects [3]. Understanding how the collapse occurring for an initially stationary granular pile, and how flow for the particles in the space-time, is an important step in efforts to manage the consequences of such natural disasters.

In past few years, a considerable amount of experiments has been carried out on the dynamic collapse of granular columns. Such laboratory tests were mainly focus on two simple configurations: cylindrical pile [4,5] or rectangular column [6–8] that was suddenly released by lifting the containing cylinder or a containing wall, respectively. Owing to the gravitational field, an avalanche event occurs and a flow is initiated, driving the column collapse toward to a pile with a typical inclination slightly lower than the repose angle [9]; the final run out distance and heap height of the granular columns is correlated to the initial aspect ratio and described by a simple power laws, which distinguish between a low and high aspect ratios; the other parameters such as the effect of size, the shape of grains, and the roughness of bottom surface were also evaluated in these works. Particularly, during the collapse process the structure of columns has been revealed that there exists a static region within which the granules are remained stationary.

To explain the dominant control parameters that govern the dynamic collapse, both the discrete element method [10–13] and the continuum approach [14] were also used to reproduce the experimental results. Compared to the properties of dynamic collapse, the quasi-static flow of granular columns such as soil creep as one of the typical examples attracts less attention despite its importance in natural phenomena and engineering application. The earliest experiment carried out by Meriaux provided a comparison with dynamic flow in the final deposits and the dissipation of energy [15]. Following this investigation, the particle stiffness, particle-wall friction, and particle-particle friction were found experimentally and numerically to be independent whereas it is very sensitive to the particle shape in the quasi-static collapse [16]. Thereafter our group observed that the slippage of single particle system is in a layered structure during the quasi-static flowing [17]. More recently, Xue et al. [14] numerically studied the quasi-static collapse of two-dimensional granular columns by using the Particle Finite Element Method, and confirmed that the collapse of columns is independent of macro friction angle and density while the container basal roughness has significant effect on the collapse. However, these previous works have only contributed to the effects of some physical factors, but lack experimental study for the dynamics of particles in quasi-static collapse process to our knowledge.

In this letter, we focus on the investigation of dynamical behaviors of particles during the quasi-static collapse process of binary granular column. Firstly, a device is constructed for moving in a controllable slow speed; Secondly, the sliding trajectory image of the particles in a two dimensional space-time is captured and analyzed by using image collection and processing technology; Thirdly, a dynamical model is proposed to describe the displacement of particles; Finally, a specially designed experiment is performed to test the calculating result of dynamical equation. Our result is helpful for the establishment of quasi-static granular flow.

## 2. Experimental

The experimental setup is schematically shown in Fig. 1. The main equipment is a rectangular container, in which the dry, cohesion less, and binary glass beads are piled. The container is enclosed on the top of an aluminum plat form by four sheets of plexus glass, including a front wall, a back wall and two side walls. The initial volume of the container is 8.0(length) × 9.0(width) × 25 (height) cm$^3$. To ensure smooth movement of the front wall and prevent the leak out of particles from the container, the gap of 0.2 cm is set up between the front wall and the other part of the container. The front wall is connected to a horizontally movable stepper motor. The velocity of stepper motor can be set to a required value. The maximum moving distance of the front wall is 8 cm. When the front wall moves at a very slow speed, the quasi-static collapse process of granular column occurs.

Under the initial container size, the binary glass beads with diameters of 0.5 cm and 0.8 cm are poured through the raining method in order: the small beads are poured first to the height 10 cm, and then the big ones are poured to the same height. The column poured is at a fixed height to facilitate homogeneity, well-reproducible and packing volume fraction. The asperities on the surface are also gently leveled obtaining a flat surface. Here, the binary granular system is used because it is more realistic than a single granular system.

After the granular column is prepared in the container, the front wall moves horizontally under the control of a stepper motor to the direction of increasing the volume of container (as depicted by the bold arrow in Fig. 1), so that it will induce a continuous collapse of the column in container. Note that, as long as the speed of wall movement remains small compared to the speed of bed relaxation, the system remains in a quasi-static regime and the flow behavior should be independent of the wall speed [14,16]. So the movement of the front wall is restricted to a very low speed (0.03 cm/s). Obviously, the difference in the movement of glass particles located at the same line perpendicular to the side wall is almost negligible during the collapse process. Therefore, the collapse profiles for the planes parallel to the side wall can be reasonably considered as the same, i.e., the collapse of the column can be uniquely described by a two-dimensional collapse profile. Consequently, the collapse rule of granular column can find out through observing the motion of particles on the side wall. For the convenience of describing, the coordinates is taken as shown in Fig. 1a in which the X and Y- axis are chosen as the moving direction of front wall and the upright direction respectively, and the origin is settled at the intersection of the back wall and the base plane.

The collapse profile is obtained by tracing the movement of particles from the transparent lateral wall. Thus, about 10% black particles of the same size and same properties were mixed into the column as tracer particles (see Fig. 1). An industrial camera (1080 × 720 pixels, 25 images/s) which is controlled by the MATLAB image acquisition program DSCM captures the positions of tracer particles. The trajectory information of a particle in the motion process is automatically tracked and processed by the Particle Tracking program according to its gray information and coordinate in the image. Due to the limited number of the black particles in one experimental measurement, it is not sufficient to catch the full picture of the collapse profile in the whole space. Hence, more than one hundred times of independent measurements are repeated to allow the tracer particles to be distributed throughout all parts of the observing space.

The sliding process of a particle can be depicted by its position and direction. Fig. 2a is the schematic diagram of trajectories of traced particles, in which some basic characteristics of the collapse can be roughly observed. The displacement of a particle varies with its location, and the closer to the upper left, the greater the displacement (see the pointing arrow). Most of the particles move toward the left lower, while those in the lower right maintain static.

In order to describe the motion behavior of all tracer particles during the collapse process, we first define a mean slipping vector. Because the trajectory change of a tracer particle can be discernible only after 5 s of quasi-static motion, we can get a displacement vector every 5 s. Therefore, a mean slipping vector can be defined as the average vector of all these displacement vectors in the whole collapse process. Next, the slipping vectors of all tracer particles are labeled in their initial positions (x, y), and then the interpolation method is used to randomly mesh the slippage (size of slipping vector) on a regular grid.

Fig. 2b is a contour of slippage on a 66 × 50 grid, where different color scales signify different slippages of particles and the black arrows represent the slipping vectors of the particles. Obviously, the distribution of slippage is a layered structure, i.e., in each sliding layer the amount of slippage is approximately the same. The layered structure found here may be associated with the layered formations of soil from the view of property of granular materials [18]. In addition, there exists a static region at the lower right part of the column in which the beads maintain repose, this means that the slippage gradually decreases from the upper left toward this static region. In addition, the gray-scale map of slippage distribution for binary granular system shows no significant difference comparing to one kind of size of particle system [17]. The reason may be that the size difference of two kinds of particles is not big enough. If the size difference is large enough, the difference of their static angles will be greater [19], resulting in a larger difference in their slippage.

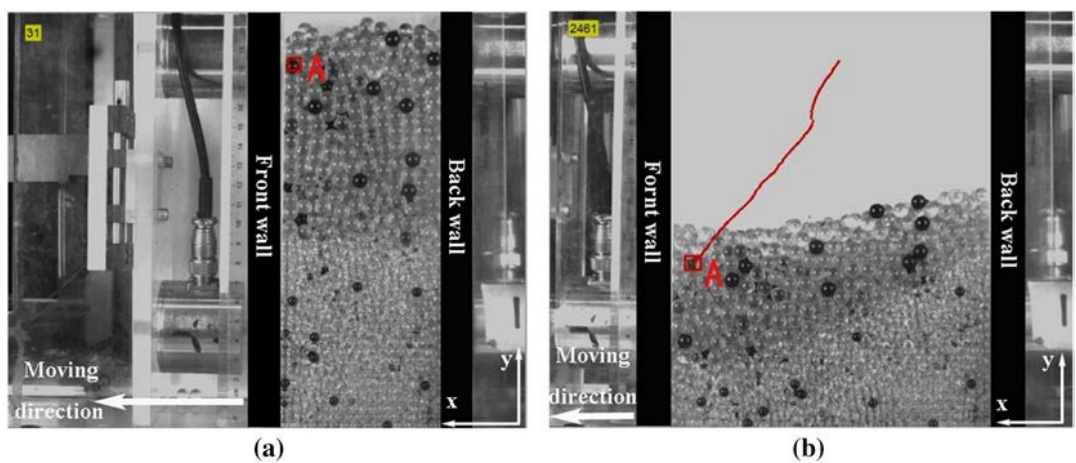

**Fig. 1.** Schematic of the experimental set up. A column of binary granular particles is laid out in a cubic container with one movable wall: (a) initial state, and (b) collapsed state. The trajectory of a large particle A is shown as the red line. (For interpretation of the references to color in this figure legend, the reader is referred to the web version of this article.)

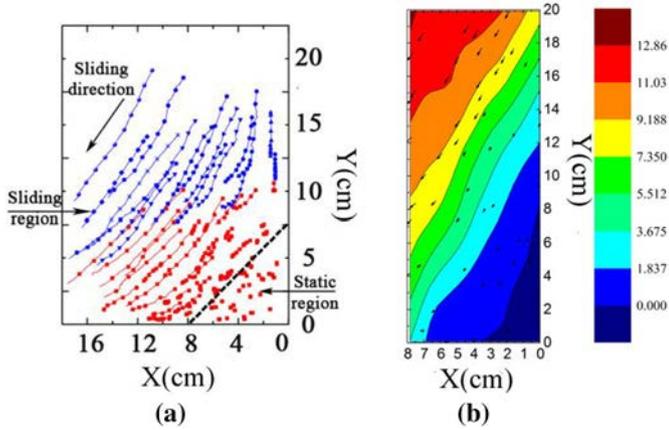

**Fig. 2.** (a) Schematic diagram of trajectories of particles during the collapse process; (b) cloud image of slippage, where different color scales signify different slippages of particles and the black arrow represents the mean slipping vector. (For interpretation of the references to color in this figure legend, the reader is referred to the web version of this article.)

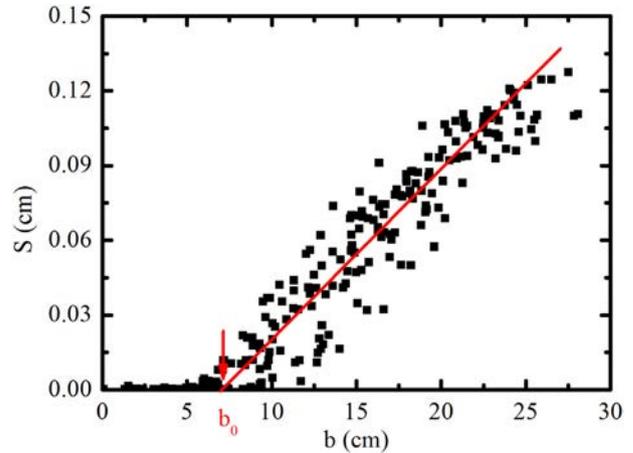

**Fig. 3.** Slippage of traced particle vs. intercept of reverse extension line of slipping vector on the Y-axis. The red line is the result of linear fitting for the slipping zones data. (For interpretation of the references to color in this figure legend, the reader is referred to the web version of this article.)

## 3. Dynamical model and experimental verification

For further investigate the inclination feature of layers, the average angle of all slipping vectors in each layer is calculated, and only the small difference exists among the average angles of all layers, i.e., all layers are almost parallel to each other. Thus, we can define a unique average angle of all slipping vectors the "layer inclination angle" $\theta_{in}$ to express the inclination of all layers relative to the X-axis:

$$\theta_{in} = \left( \sum_{l=1}^{n} \arctan[S_{ly}/S_{lx}] \right)/n \quad (1)$$

where $n$ and $l$ are the total number and the sequence number of the traced particles, $S_{ly}$ and $S_{lx}$ are the vertical and horizontal components of slipping vector **S**, respectively. According to Eq. (1), the inclination angles of large particle layers and small particle layers are calculated to be 54° and 45° respectively. Considering the calculation error however, we cannot be sure whether this angle difference is true or not, so we take an average value 49.5° in the following calculation. This average slipping angle is slightly different from that of a single particle system [17].

Next, to reveal the motion state of a particle, we first use a linear equation to determine the position of the slipping layer that the particle belongs to:

$$y = kx + b \quad (2)$$

where $k = -\tan(\theta_{in})$ represents the slop of layers, and $b$ is an average value of intercept of the layer boundary on Y-axis. Different $b$ values corresponds different positions of layers. Furthermore, we can explore the relationship between the slippage of a particle and the position of layer in which it exists. Taking the Cartesian coordinate of a traced particle into Eq. (2), $b$ value is obtained, and meanwhile the slippage S is deduced. Fig. 3 is the relationship between the slippages of traced particles and the intercepts of reverse extension line of their slipping vectors on the Y-axis. It is easy to find that, there exists a threshold value $b_0$ below which the slippages approximately equal to zero, i.e., no slipping occurs in this region, while above which the slippages increases almost linearly with the position of layer. This means that there exists a boundary layer which separates between static and sliding zones. The calculated $b_0$ is about 7 cm in this system, which is one-third of the initial height of the column.

To gain insight into the dynamics of particles in quasi-static collapse process, we establish a simple theoretical model to predict the trajectory evolution of particles in the space-time. Generally, the sliding process of a particle should be described by its position and direction at any time. Now, the motion direction of a particle can be directly determined by the average slipping angle $\theta_{in}$. Next, if the initial position coordinate of the particle is known, the layer in which the particle exists can be determined by calculating $b$ of Eq. (2), and the slippage of this particle can be known further based on the date of Fig. 2b. Taking into account the quasi-static and uniform flow of particles, the average velocity $v$ of the particle can be calculated through the sliding distance divided by the total moving time. Then, the positions of a particle in the collapse process can be written as the following.

$$\begin{cases} x_i = x_0 + v\cos(\theta_{in})\Delta t \\ y_i = y_0 + v\cos(\theta_{in})\Delta t \end{cases} \quad (3)$$

where $(x_i,y_i)$ is the coordinate at any time, $(x_0,y_0)$ is the coordinate at initial time, $\Delta t$ is the moving time.

In order to verify the correctness of dynamic Eq. (3), an experiment is especially devised: two sizes of particles were dyed in two colors, and the granular column is dyed into 4 layers. As shown in Fig. 4, three easily observed boundaries can be formed. The trajectory evolution of particles located at three boundary lines not only can be observed in the process of collapse, but also can be quantitatively calculated by Eq. (3). The black lines correspond to the result of calculating. By comparing the real moving trajectories of particles with the theoretical calculation shown as the black lines, it can be seen that the model proposed can predict the dynamical behavior of particles. The calculating result for the trajectories of particles on the selected layers is well consistent with the experimental observation.

## 4. Conclusion

In conclusion, an experimental setup is developed to control the collapse process of a binary granular system quasi-statically. By using the particle tracking technique and gray analysis method, we find that the collapse can be approximately described by the layers sliding. Moreover, there exists a boundary-layer which separates between the static and the sliding zones, below which no sliding happens. The dynamic model established here can quantitatively explain the sliding angle and distance for any particle in the quasi-static collapse process.

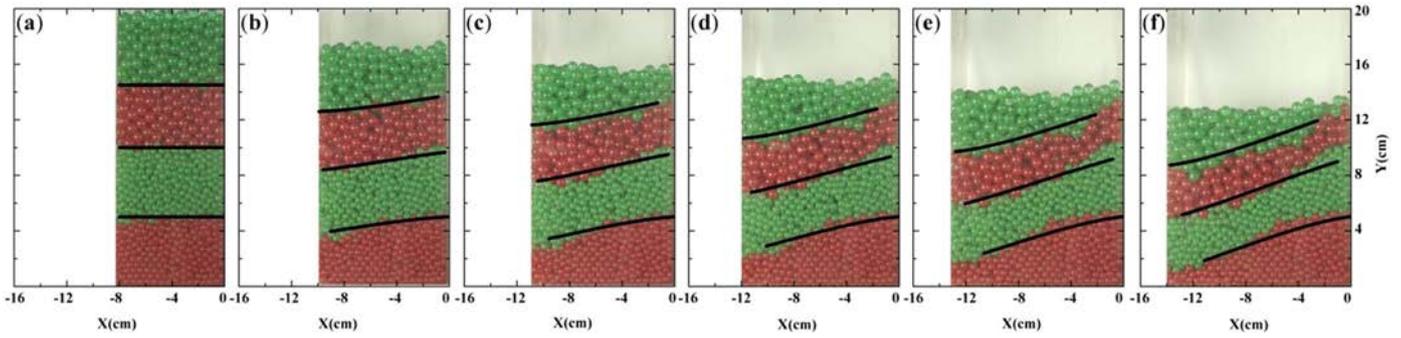

**Fig. 4.** Photographs from (a) to (f) show the trajectory evolution of particles at time series of 0 min, 1 min, 1.5 min, 2 min, 2.5 min and 3 min, respectively.

However, it should be noted that we only considered the spherical beds as research object in this study. In fact, the shape of particles, the friction coefficient of particles, and the size ratio of binary particles may affect the dynamical behavior of particles in the process of quasi-static collapse. These possible effects of such factors should be explored by means of combining experiment with DEM method [20,21] in future work. Moreover, the comparison between the dynamical model and the experimental result in Fig. 4 is qualitative. In fact, if the quantitative comparison is carried out, the tracer particle technology can still be used. That is, the tracer particles are placed in different positions on the boundary line, so that the position coordinates of the corresponding particles at different times of the collapse process can be observed. After repeating the experiments for many times, take the averages of these position coordinates at different times as the evolutionary trajectory. Then the position coordinates of these tracer particles at different times are compared with those calculated according to Eq. (3), meanwhile the relative error can be given. Finally, using transparent soil for describing the catastrophic collapse of natural materials (i.e., soils, rocks, geomaterials, etc.) is more practical than glass beads though it is very challenging.

### Acknowledgements

This work was supported by the Chinese National Science Foundation, Project Nos. 10975014, 11274355 and 11475018.

### References

[1] R.M. Iverson, The physics of debris flows, Rev. Geophys. 35 (1997) 245–296.
[2] D.R. Scott, Seismicity and stress rotation in a granular model of the brittle crust, Nature 381 (1996) 592–595.
[3] Associated Press, Silo Collapse Sends about 10,000 Tons of Corn onto Road, http://www.nydailynews.com/news/national/silo-collapse-spills-10-000-tons-corn-road-article-1.3772569?barcprox=true 2018, January 22.
[4] T.S. Komatsu, S. Inagaki, N. Nakagawa, S. Nasuno, Creep motion in a granular pile exhibiting steady surface flow, Phys. Rev. Lett. 86 (2001) 1757–1760.
[5] P.-A. Lemieux, D.J. Durian, From avalanches to fluid flow: a continuous picture of grain dynamics down a heap, Phys. Rev. Lett. 85 (2000) 4273–4276.
[6] D. Fenistein, J.W. Meent, M.V. Hecke, Universal and wide shear zones in granular bulk flow, Phys. Rev. Lett. 92 (2004) 094301–094304.
[7] I.I. Albert, P. Tegzes, B. Kahng, R. Albert, J.G. Sample, M. Pfeifer, A. Barabasi, T. Vicsek, P. Schiffer, Jamming and fluctuations in granular drag, Phys. Rev. Lett. 84 (2000) 5122–5125.
[8] C. Mériaux, Two dimensional fall of granular columns controlled by slow horizontal withdrawal of a retaining wall, Phys. Fluids 18 (9) (2006) 399–486.
[9] J. Lee, H.J. Herrmann, Angle of repose and angle of marginal stability: molecular dyanmics of granular particles, J. Phys. A 26 (2) (1993) 373.
[10] F. Bertrand, L.A. Leclaire, G. Levecque, DEM-based models for the mixing of granular materials, Chem. Eng. Sci. 60 (8) (2005) 2517–2531.
[11] S. Utili, R. Nova, DEM analysis of bonded granular geomaterials, Int. J. Numer. Anal. Methods Geomech. 32 (17) (2008) 1997–2031.
[12] C. Kloss, C. Goniva, A. Hager, S. Amberger, S. Pirker, Models, algorithms and validation for open source DEM and CFD–DEM, Prog. Comput. Fluid Dyn. Int. J. 12 (2) (2012) 140–152.
[13] K.E. N'Tsoukpoe, G. Restuccia, T. Schmidt, X. Py, The size of sorbents in low pressure sorption or thermochemical energy storage processes, Energy 77 (2014) 983–998.
[14] X. Zhang, Y. Ding, D. Sheng, S.W. Sloan, W. Huang, Quasi-static collapse of two dimensional granular columns: insight from continuum modeling, Granul. Matter 18 (3) (2016) 1–14.
[15] Catherine Mériaux, Two dimensional fall of granular columns controlled by slow horizontal withdrawal of a retaining wall, Phys. Fluids 18 (9) (2006) 399–486.
[16] P.J. Owen, P.W. Cleary, C. Mériaux, Quasi-static fall of planar granular columns: comparison of 2D and 3D discrete element modeling with laboratory experiments, Geomech. Geoeng. 4 (1) (2009) 55–77.
[17] B.C. Pan, Q.F. Shi, G. Sun, Experimental observation of quasi-static avalanche process of a granular pile, Chin. Phys. Lett. 30 (12) (2013), 124701. .
[18] J.K. Mitchell, K. Soga, Fundamentals of Soil Behavior, 3rd ed. John Wiley & Sons, Inc, New Jersey, 2005https://doi.org/10.2136/sssaj1976.03615995004000040003x.
[19] Z. Liu, Measuring the Angle of Repose of Granular Systems Using Hollow Cylinders, University of Pittsburgh, 2011 http://d-scholarship.pitt.edu/6401/ (MS Thesis).
[20] H.M.B. Al-Hashemi, O.S.B. Al-Amoudi, A review on the angle of repose of granular materials, Powder Technol. 330 (2018) 397–417.
[21] H.T. Chou, C.F. Lee, Y.C. Chung, S.S. Hsiau, Discrete element modeling and experimental validation for the falling process of dry granular step, Powder Technol. 231 (2012) 122–134.